\documentclass[%
nofootinbib,
 amsmath,amssymb,
 aps,
 prd,
twocolumn, 
superscriptaddress
]{revtex4-1}

\usepackage[normalem]{ulem}
\usepackage{cancel}

\usepackage{graphicx}
\usepackage{dcolumn}
\usepackage{bm}
\usepackage{orcidlink}
\hypersetup{colorlinks, citecolor=teal}
\hypersetup{colorlinks=true, linkcolor=blue, urlcolor=purple}
\usepackage{hyperref}
\usepackage{footnote}

\usepackage{xcolor}
\definecolor{darkblue}{rgb}{0.2, 0.2, 0.6}

\begin{document}


\title{Moment of inertia and compactness of quark stars within the context of $f(R,T,L_m)$ gravity}

\author{Juan M. Z. Pretel \orcidlink{0000-0003-0883-3851}}
 \email{juanzarate@cbpf.br}
 \affiliation{
 Centro Brasileiro de Pesquisas F{\'i}sicas, Rua Dr.~Xavier Sigaud, 150 URCA, Rio de Janeiro CEP 22290-180, RJ, Brazil
}

\date{\today}

\begin{abstract}
Within the metric formalism of $f(R,T,L_m)$ gravity theories, we investigate the hydrostatic equilibrium structure of compact stars taking into account both isotropic and anisotropic pressure. For this purpose, we focus on the $f(R,T,L_m)= R+ \alpha TL_m$ model, where $\alpha$ is a free parameter. We derive the modified TOV equations and the relativistic moment of inertia in the slowly rotating approximation. Using an equation of state (EoS) for color-superconducting quark stars, we examine the effects of the $\alpha TL_m$ term on the different macroscopic properties of these stars. Our results reveal that the decrease of the parameter $\alpha$ leads to a noticeable increase in the maximum-mass values. For negative $\alpha$ with sufficiently small $\vert\alpha\vert$, we obtain a qualitative behavior similar to the general relativistic (GR) context, namely, it is possible to obtain a critical stellar configuration such that the mass reaches its maximum. However, for sufficiently large values of $\vert\alpha\vert$ keeping negative $\alpha$, the critical point cannot be found on the mass-radius diagram. We also find that the inclusion of anisotropic pressure can provide masses and radii quite consistent with the current observational measurements, which opens an outstanding window onto the physics of anisotropic quark stars. By comparing the $I-C$ relations of isotropic quark stars in modified gravity, we show that such a correlation remains almost unchanged as the parameter $\alpha$ varies from GR counterpart. On the other hand, given a fixed $\alpha$, the $I-C$ relation is insensitive to variations of the anisotropy parameter $\beta$ to $\mathcal{O}(4\%)$.

\end{abstract}

\maketitle

\section{Introduction}

Despite the great success of General Relativity (GR) in predicting a series of gravitational phenomena ranging from the solar system \cite{Will2014} to the emission of gravitational waves emanating from compact-object mergers \cite{Abbott221101, Abbott011102, Abbott122002, Kyutoku2021}, the $\Lambda$CDM model faces theoretical and observational challenges which suggest that we have to look beyond GR as the underlying gravity theory \cite{Shankaranarayanan2022}. In this standard model of modern cosmology, the most prominent description of dark energy is the cosmological constant $\Lambda$ which fits very well with the observational data but suffers from the well-known coincidence problem and the fine-tuning problem \cite{Weinberg1989, Padmanabhan2003}. Nevertheless, it is possible to obtain a cosmic expansion scenario of the universe through modified gravity theories without the need to introduce exotic forms of fluid such as dark energy \cite{Starobinsky1980, Boisseau2000, Esposito2001, Capozziello2002, Carroll2004, Copeland2006, Hu2007, Nojiri2007, Amendola2007, Appleby2007, Koyama2016, Odintsov2020}. One of the simplest ways to modify GR is by replacing the Ricci scalar $R$ in the usual Einstein-Hilbert action by a modified Lagrangian $f(R)$, namely, the so-called $f(R)$ gravity theories \cite{Sotiriou2010, DeFelice2010}. These theories have been intensively investigated in the last two decades in order to explain the evolution of the early and present universe, see e.g.~the review articles \cite{Capozziello2011, Nojiri2011, Clifton2012, Nojiri2017} for a broader understanding on the subject. Moreover, at a smaller (astrophysical) level, $f(R)$ theories have also been tested using compact stars \cite{Zubair2016, Folomeev2018, Astashenok2020, Astashenok2021, AstashenokEPL, Nashed2021EPJC, Silveira2021, Jimenez2022, Numajiri2022, Pretel2022CQG, Pretel2022JCAP, Bora2022JCAP, Astashenok2022, Oikonomou2023MNRAS, Numajiri2023, Nashed2024EPJC}.

Bertolami et al.~\cite{Bertolami2007} proposed an extension of $f(R)$ gravity including an explicit coupling between an arbitrary function of the scalar curvature and the matter Lagrangian density $L_m$. Interestingly, as a result of the coupling, they found that the equation of motion of massive particles is non-geodesic, and an extra force, orthogonal to the four-velocity, arises. Years later, Harko and Lobo \cite{HarkoLobo2010} suggested a more evolved generalization of matter-curvature coupling theories, thus giving rise to the so-called $f(R,L_m)$ gravity theories. The energy conditions \cite{Wang2012}, thermodynamics \cite{Pourhassan2020}, cosmological models \cite{Kavya2022, Jaybhaye2022, Maurya2023, Myrzakulova2024}, compact stars \cite{Carvalho2020EPJC, Lobato2021EPJC, Carvalho2022EPJC} and traversable wormhole solutions \cite{Venkatesha2023, Kavya2023, Naseer2023} have been recently studied within the framework of $f(R,L_m)$ gravity.

On the other hand, Harko and collaborators \cite{Harko2011} proposed another extension of standard GR, the $f(R,T)$ modified gravity theories, where the gravitational Lagrangian is given by an arbitrary function of the Ricci scalar $R$ and of the trace of the energy-momentum tensor $T$. The cosmological consequences \cite{Shabani2018, Debnath2019, Bhattacharjee2020, Bhattacharjee2020EPJP, Baffou2021, Gamonal2021, Chen2022, Singh2023, Sarkar2023, Mohammadi2023, Sofuoglu2023, Jeakel2024} and applications to compact stars \cite{Deb2018, Deb2019MNRAS, Maurya2019PRD, Deb2019JCAP, Pretel2021JCAPa, Pretel2021JCAPb, Biswas2021AoP, PretelCPC, Bora2022, Pretel2022MPL, Pappas2022, Nashed2023, Nashed2023EPJC, Bhattacharya2024} of such a theory have been addressed by several authors in recent years. See further Refs.~\cite{Goncalves2022PRD, Pinto2022, Goncalves2022, Bouali2023} for a novel scalar-tensor representation of $f(R,T)$ gravity. A generalized and unified formulation of $f(R,T)$ and $f(R, L_m)$ type gravity models consists of considering the gravitational Lagrangian density as an arbitrary function of $R$, $T$ and $L_m$, namely, the $f(R,T,L_m)$ theories. The field equations for this maximal extension of the Hilbert–Einstein action were recently derived from a variational principle by Haghani and Harko \cite{Haghani2021}, where they have examined the cosmological implications for a homogeneous, isotropic and flat universe by considering the simplest model, i.e.~the $f(R,T,L_m)= R+ \alpha TL_m$ gravity model with $\alpha$ being a free parameter of the theory. Recently, this theory has been tested at scales typical of compact stars \cite{Mota2024arxiv, Tangphati2024arxiv}, where the impact of $\alpha$ on the mass-radius diagrams has been investigated. The present study aims to calculate the relativistic moment of inertia $I$ and investigate its relation with the compactness $C$ (i.e., the $I-C$ relation), and hence generalize previous works. As we will mention below, this correlation can be very useful when determining the radius of a compact star.

On one hand, different modified gravity theories have a significant effect on the basic global characteristics of a compact star \cite{Olmo2020}. On the other hand, it has also been shown that anisotropic pressure plays an important role in describing self-gravitating systems at very high energy densities \cite{BowersLiang1974, HerreraSantos1997, Horvat2011, Doneva2012, Yagi2015, Pretel2020EPJC, Kumar2022}. In turn, different anisotropy models are capable of producing results consistent with observational measurements \cite{Biswas2019, Rahmansyah2020, Rahmansyah2021, Pretel2024PLB, Das2023}. Of course, under certain limits, one must recover the isotropic solutions in any gravity theory. So our main task in this study is to examine the effect of the parameter $\alpha$ on the compact star structure and analyze the deviations with respect to the GR results under the presence of anisotropic pressure. It is also worth pointing out that gravitational-wave events can be used to test gravity theories in the strong-field regime, as well as to constraint the EoS employed to describe matter at nuclear densities. In fact, it has been shown that the secondary component of the GW190814 event \cite{Abbott2020} can be consistently described as a neutron star (NS) in the context of $R$-squared gravity \cite{Astashenok2020, Astashenok2021, AstashenokEPL}. With this in mind, we will compare our theoretical results with recent observational mass-radius measurements.

Within the gravitational framework of GR, it has been shown that it is possible to express various global properties of compact stars in terms of empirical functions (usually referred to as ``universal relations'') that do not depend on the specific EoS. Particularly, Ravenhall and Pethick \cite{Ravenhall1994} emphasized that the dimensionless moment of inertia $I/(Mr_{\rm sur}^2)$ and stellar compactness $C= M/r_{\rm sur}$ are connected by an apparently universal relation, where $r_{\rm sur}$ and $M$ are the radius and total mass of the star, respectively. Although a variety of EoSs lead to considerably different $M-r_{\rm sur}$ relations, the authors showed for the first time that the normalized moment of inertia as a function of $C$ is surprisingly similar. These universal relations, given as low-order polynomial expressions, were since refined by other researchers \cite{Lattimer2001, Bejger2002}. Following such a procedure, Bejger and Haensel \cite{Bejger2002} derived certain limits for the Crab pulsar. Furthermore, using a higher-order polynomial function \cite{Lattimer2005}, it was argued that it is possible to determine the radius of a NS from the $I-C$ relation once $I$ and $M$ are measured, as in the case of a pulsar within a binary system. Thus, the simultaneous mass and moment of inertia measurements can constrain the radius of a compact star. Motivated by these works, we aim to investigate whether the $I-C$ relation holds in $f(R,T,L_m)$ modified gravity theories. It is important to remark that Breu and Rezzolla \cite{Breu2016} showed that the universal behavior of the dimensionless moment of inertia $I/M^3$ is more accurate than that used in previous studies. In this work we will therefore use $I/M^3$ instead of $I/(Mr_{\rm sur}^2)$.

The rest of the paper is organized as follows: In Sect.~\ref{Sec2} we briefly summarize $f(R,T,L_m)$ gravity and present the field equations for the specific function $f(R,T,L_m)= R+ \alpha TL_m$. In Sect.~\ref{Sec3} we derive the stellar structure equations for an anisotropic fluid sphere in a state of hydrostatic equilibrium. To determine the relativistic moment of inertia of slowly rotating anisotropic quark stars (QSs) up to first order in the angular velocity, we follow a procedure similar to that carried out by Hartle in GR \cite{Hartle1967}, as described in the third part of Sect.~\ref{Sec3}. For our numerical calculations, we use a physically relevant EoS for the radial pressure and a phenomenological ansatz for the anisotropy profile (known in the literature as quasi-local model \cite{Horvat2011}), where a parameter $\beta$ quantifies the anisotropy within the stellar fluid. These two inputs for the modified Tolman-Oppenheimer-Volkoff (TOV) equations are presented in Sect.~\ref{Sec4}. Then the section \ref{ResultsSec} is devoted to the numerical results for isotropic and anisotropic QSs, and we report their global properties in terms of the free parameters $\alpha$ and $\beta$. Finally, in Sect.~\ref{Conclusions} we provide our conclusions.


\section{\bf Basic formalism of $f(R,T, L_m)$ gravity}\label{Sec2}

A generalized and unified formulation of $f(R,T)$ and $f(R, L_m)$ type gravity models consists of considering the gravitational Lagrangian density as an arbitrary function of the Ricci scalar $R$, of the trace of the energy-momentum tensor $T$, and of the matter Lagrangian $L_m$, so that $L_g= f(R,T,L_m)$ \cite{Haghani2021}. Accordingly, the modified Einstein-Hilbert action is given by
\begin{equation}\label{AcionEq}
    S = S_g + S_m = \int\sqrt{-g} \left[ \frac{1}{16\pi}f(R,T,L_m) + L_m \right] d^4x ,
\end{equation}
where $g$ is the determinant of the metric tensor $g_{\mu\nu}$. The variation of the gravitational action $S_g$ with respect to the inverse metric leads to 
\begin{align}\label{DeltaSgEq}
    \delta S_g =&\ \frac{1}{16\pi} \int\sqrt{-g} \bigg[ f_RR_{\mu\nu} - \left( \frac{f}{2}- f_mL_m \right)g_{\mu\nu}   \nonumber \\
    &\ + (g_{\mu\nu}\square - \nabla_\mu\nabla_\nu)f_R  - f_mT_{\mu\nu} - f_T\tau_{\mu\nu}\bigg] \delta g^{\mu\nu}d^4x ,
\end{align}
with $R_{\mu\nu}$ being the Ricci curvature tensor
and $f_m= f_T+ f_L/2$, where we have denoted $f_R= \partial f/\partial R$, $f_T= \partial f/\partial T$, $f_L= \partial f/\partial L_m$. Moreover, $\square = \nabla_\mu\nabla^\mu$ is the d'Alembert operator with $\nabla_\mu$ standing for the covariant derivative associated with the Levi-Civita connection, and the new tensor $\tau_{\mu\nu}$ is defined as
\begin{equation}\label{tauEq}
    \tau_{\mu\nu} = 2g^{\alpha\beta}\frac{\partial^2L_m}{\partial g^{\mu\nu}\partial g^{\alpha\beta}} , 
\end{equation}
while the ordinary (baryonic) matter energy-momentum tensor can be written as 
\begin{equation}
    T_{\mu\nu} = \frac{-2}{\sqrt{-g}}\frac{\delta S_m}{\delta g^{\mu\nu}} = g_{\mu\nu}L_m- 2\frac{\partial L_m}{\partial g^{\mu\nu}} .
\end{equation}

From the variational principle $\delta S=0$ for the action (\ref{AcionEq}) and in view of Eq.~(\ref{DeltaSgEq}), we hence obtain the following field equations 
\begin{align}\label{FieldEq}
    f_RR_{\mu\nu} &- \left( \frac{f}{2}- f_mL_m \right)g_{\mu\nu} + (g_{\mu\nu}\square - \nabla_\mu\nabla_\nu)f_R  \nonumber  \\
    &= \left( 8\pi+ f_m \right)T_{\mu\nu} + f_T\tau_{\mu\nu} .
\end{align}
Note that Eq.~(\ref{FieldEq}) reduces to the conventional Einstein equation when $f(R,T,L_m) =R$. Furthermore, the trace of the field equations (\ref{FieldEq}) leads to
\begin{align}\label{TraceEq}
    3\square f_R+ Rf_R - 2(f- 2f_mL_m) = (8\pi + f_m)T + f_T\tau .
\end{align}

By taking into account the geometric identities $(\square\nabla_\nu - \nabla_\nu\square)\phi = R_{\mu\nu}\nabla^\mu\phi$ and $\nabla_\mu R^{\mu\nu} = \nabla^\nu R/2$, the covariant divergence of the field equations (\ref{FieldEq}) gives rise to the following non-conservative equation for the energy-momentum tensor
\begin{align}\label{NonConservEq}
    \nabla^\mu T_{\mu\nu} =&\ \frac{1}{8\pi+ f_m}\bigg[ \nabla_\nu(f_mL_m) - T_{\mu\nu}\nabla^\mu f_m  \nonumber  \\
    &- \frac{1}{2}(f_T\nabla_\nu T + f_L\nabla_\nu L_m) - \nabla^\mu(f_T\tau_{\mu\nu}) \bigg] .
\end{align}

We are interested in obtaining numerical solutions describing the hydrostatic equilibrium of anisotropic compact stars, so it becomes necessary to specify a particular function for the $f(R,T,L_m)$ gravity theory. The simplest model involving a minimal matter-gravity coupling in such theories is given by $f(R,T,L_m)= R+ \alpha TL_m$, whose cosmological implications have been investigated for a homogeneous, isotropic and flat universe in Ref.~\cite{Haghani2021}. The constant parameter $\alpha$ has dimensions $[\rm length]^2$ (in a geometric unit system), is the only free parameter of the gravitational theory, and the specific case $\alpha= 0$ corresponds to pure Einstein gravity. Under such assumption, we have $f_m= \alpha(T+ 2L_m)/2$, and Eqs.~(\ref{FieldEq}), (\ref{TraceEq}) and (\ref{NonConservEq}) can be written, respectively, as follows
\begin{align}
    G_{\mu\nu} =& \left[ 8\pi+ \frac{\alpha}{2}(T+ 2L_m) \right]T_{\mu\nu} + \alpha L_m(\tau_{\mu\nu}- L_mg_{\mu\nu}) ,  \label{FieldEq1} 
\end{align}
\begin{equation}
    R = - \left[ 8\pi+ \frac{\alpha}{2}(T+ 2L_m) \right]T - \alpha L_m(\tau- 4L_m) ,  \label{TraceEq1}
\end{equation}
\begin{align}
    \nabla^\mu T_{\mu\nu} =&\ \frac{2\alpha}{16\pi+ \alpha(T+ 2L_m)}\left[ \nabla_\nu\left( L_m^2+ \frac{1}{2}TL_m \right) \right.  \nonumber  \\
    &- \frac{1}{2}T_{\mu\nu}\nabla^\mu(T+ 2L_m) -\nabla^\mu(L_m\tau_{\mu\nu}) \nonumber  \\
    &\left. - \frac{1}{2}(L_m\nabla_\nu T + T\nabla_\nu L_m) \right] ,  \label{NonConservEq1}
\end{align}
where $G_{\mu\nu}$ is the standard Einstein tensor. When $\alpha= 0$, Eq.~(\ref{FieldEq1}) reduces to the usual Einstein equation $G_{\mu\nu}= 8\pi T_{\mu\nu}$. From Eq.~(\ref{NonConservEq1}), it is evident that the energy-momentum tensor is a non-conserved quantity. Notwithstanding, the fact that the four-divergence of the energy–momentum tensor is non-zero opens the possibility of a gravitationally induced particle production as shown by Harko and colleagues \cite{Harko2014, Harko2015EPJC}. In fact, there exists a considerable number of modified gravity theories allowing departures from the usual conservative framework. For a review of non-conservative gravity theories, we refer the reader to Ref.~\cite{Velten2021}.


\section{Stellar structure equations}\label{Sec3}

\subsection{Modified TOV equations}

As in GR, we consider a 4-dimensional Lorentz manifold with metric $g_{\mu\nu}$ of signature $(-,+,+,+)$. The equilibrium anisotropic configuration is assumed to be static and spherically symmetric, so that the spacetime geometry can be described by the familiar line element 
\begin{equation}\label{MetricEq}
    ds^2 = -e^{2\psi}dt^2 + e^{2\lambda}dr^2 + r^2(d\theta^2 + \sin^2\theta d\phi^2) ,
\end{equation}
where the metric functions $\psi$ and $\lambda$ depend only on the radial coordinate and $x^\mu = (t,r,\theta,\phi)$ is the 4-position vector. We are interested in investigating the astrophysical effects of the $\alpha TL_m$ term on the relativistic structure of a compact star made of anisotropic matter. Realistic models of compact stars must be constructed taking into account anisotropic pressures since anisotropy could manifest itself at high densities \cite{HerreraSantos1997, Mak2003, Horvat2011, Kumar2022}. Keeping this in mind, we conjecture that the matter-energy content of the star can be represented by an anisotropic perfect fluid, with energy-momentum tensor in its covariant form written as 
\begin{equation}\label{EMTensor}
    T_{\mu\nu} = (\rho+ p_t)u_\mu u_\nu + p_tg_{\mu\nu} - \sigma k_\mu k_\nu ,
\end{equation}
where $\rho$ is the energy density, $p_t$ the transverse pressure and $\sigma= p_t-p_r$ is the anisotropy factor with $p_r$ being the radial pressure. Moreover, $u^\mu$ is the four-velocity of a comoving fluid element in the source, $k^\mu$ is a unit spacelike four-vector, and they must satisfy the relations: $u_\mu u^\mu= -1$, $k_\mu k^\mu= 1$, and $u_\mu k^\mu= 0$. Evidently, Eq.~(\ref{EMTensor}) corresponds to an isotropic fluid when $\sigma$ vanishes. In view of the metric (\ref{MetricEq}), we have $u^\mu= e^{-\psi}\delta_0^\mu$ and $k^\mu= e^{-\lambda}\delta_1^\mu$. The trace of the energy-momentum tensor (\ref{EMTensor}) is simply given by $T= g^{\mu\nu}T_{\mu\nu}= -\rho+ 3p_r+ 2\sigma$.

Since there is no unique definition of the matter Lagrangian density, in the present study we adopt the expression $L_m = \mathcal{P}$, where $\mathcal{P}= (p_r+2p_t)/3$ for anisotropic matter \cite{Deb2019MNRAS, Maurya2019PRD, Biswas2021AoP, Pretel2022MPL}, so that $\tau_{\mu\nu}= 0$. With this assumption, Eq.~(\ref{FieldEq1}) becomes 
\begin{equation}\label{FieldEqPartForm}
    G_{\mu\nu} = \left[ 8\pi+ \frac{\alpha}{2}\left( -\rho+ 5p_r +\frac{10}{3}\sigma \right) \right]T_{\mu\nu} - \alpha \mathcal{P}^2g_{\mu\nu} ,
\end{equation}
whose non-vanishing components are given explicitly by 
\begin{align}\label{FieldEq00}
    \frac{1}{r^2}\frac{d}{dr}\left( re^{-2\lambda} \right) - \frac{1}{r^2} =& -\left[ 8\pi+ \frac{\alpha}{2}\left( 5p_r- \rho+ \frac{10}{3}\sigma \right) \right]\rho  \nonumber  \\
    &- \alpha\left( p_r+ \frac{2}{3}\sigma \right)^2 ,
\end{align}
\begin{align}\label{FieldEq11}
    e^{-2\lambda}\left( \frac{2}{r}\psi'+ \frac{1}{r^2} \right) - \frac{1}{r^2} =& \left[ 8\pi+ \frac{\alpha}{2}\left( 5p_r- \rho+ \frac{10}{3}\sigma \right) \right]p_r  \nonumber  \\
    &- \alpha\left( p_r+ \frac{2}{3}\sigma \right)^2 ,
\end{align}
\begin{align}\label{FielEq22}
    &e^{-2\lambda}\left[ \psi''+ \psi'^2 - \psi'\lambda' + \frac{1}{r}(\psi' - \lambda') \right]  \nonumber  \\
    =& \left[ 8\pi+ \frac{\alpha}{2}\left( 5p_r- \rho+ \frac{10}{3}\sigma \right) \right](p_r+ \sigma) - \alpha\left( p_r+ \frac{2}{3}\sigma \right)^2 ,
\end{align}
where the prime represents differentiation with respect to $r$. The idea is to algebraically manipulate these field equations in order to obtain an equation describing the mass distribution, and the hydrostatic equilibrium equation. Furthermore, the non-conservation equation (\ref{NonConservEq1}) becomes 
\begin{equation}
    \nabla^\mu T_{\mu\nu} = \frac{\alpha/2}{8\pi+ \frac{\alpha}{2}(2\mathcal{P}+ T)} \bigg[ 4\mathcal{P} \partial_\nu\mathcal{P} - T_\nu^\mu \partial_\mu(2\mathcal{P} + T) \bigg] ,
\end{equation}
which, for the index $\nu= 1$, leads to the following expression 
\begin{align}\label{PressPrime}
    p_r' =& -(\rho+ p_r) \psi' + \frac{2}{r}\sigma  \nonumber \\
    &+ \frac{\alpha\left[ p_r(\rho'-p_r'-2\sigma'/3) + 8\sigma(p_r'+ 2\sigma'/3)/3 \right]}{16\pi+ \alpha(5p_r- \rho+ 10\sigma/3)} .
\end{align}

From the above equations (\ref{FieldEq00}) and (\ref{FieldEq11}) we can obtain respectively:
\begin{align}\label{MassFunc}
    m(r) =&\ 4\pi\int_0^r \bar{r}^2\rho(\bar{r})d\bar{r} + \frac{\alpha}{2}\int_0^r \bar{r}^2 \bigg[ \frac{\rho(\bar{r})}{2}\bigg( 5p_r(\bar{r})- \rho(\bar{r})  \nonumber  \\
    &+ \frac{10}{3}\sigma(\bar{r}) \bigg)+ \mathcal{P}^2(\bar{r}) \bigg]d\bar{r} ,
\end{align}
and
\begin{align}\label{PsiPrime}
    \psi' =& \left[ \frac{m}{r^2}+ 4\pi rp_r + \frac{\alpha}{4}\left(  3p_r -\rho+ \frac{2}{3}\sigma \right)rp_r- \frac{2\alpha}{9}r\sigma^2 \right]  \nonumber  \\
    &\times \left( 1- \frac{2m}{r} \right)^{-1} ,
\end{align}
where the metric potential $\lambda(r)$ is determined via the relation
\begin{equation}\label{ExpLambda}
    e^{-2\lambda} = 1 - \frac{2m}{r} .
\end{equation}
\vspace{0.01cm}

Note that, in Eq.~(\ref{MassFunc}), $\bar{r}$ is the integration variable, and has been denoted that way in order not to confuse it with the upper limit of the integral. The mass function $m(r)$ in such equation represents nothing more than the gravitational mass within a sphere of radius $r$. The first integral is the widely known expression in the pure general relativistic scenario, while the second integral comes to be an extra mass contribution due to the $\alpha TL_m$ term. Part of this work is to investigate how the mass of a compact star is affected by the presence of this contribution. At the surface $r= r_{\rm sur}$, where the radial pressure drops to zero, the total mass of the anisotropic compact star is given by $M = m(r_{\rm sur})$. Particularly, we will investigate the effect of the parameter $\alpha$ on the $M-r_{\rm sur}$ curves taking into account the two cases $\sigma= 0$ and $\sigma\neq 0$.

Using the Eqs.~(\ref{PressPrime})-(\ref{PsiPrime}), we can explicitly write down the modified TOV equations for the relativistic structure of an anisotropic compact star within the framework of $f(R,T,L_m)= R+ \alpha TL_m$ gravity: 
\begin{widetext}
\begin{subequations}
  \begin{align}
    \frac{dm}{dr} =&\ 4\pi r^2\rho + \frac{\alpha}{2}r^2\left[ \frac{\rho}{2}\left( 5p_r- \rho+ \frac{10}{3}\sigma \right) + \left( p_r+ \frac{2}{3}\sigma \right)^2 \right] ,  \label{TOV1}  \\
    \frac{dp_r}{dr} =& -(\rho+ p_r)\left[ \frac{m}{r^2}+ 4\pi rp_r + \frac{\alpha}{4}\left(  3p_r -\rho+ \frac{2}{3}\sigma \right)rp_r- \frac{2\alpha}{9}r\sigma^2 \right]\left( 1-\frac{2m}{r} \right)^{-1} + \frac{2}{r}\sigma  \nonumber \\ 
    &+ \frac{\alpha}{16\pi+ \alpha(5p_r- \rho+ 10\sigma/3)}\left[ p_r\left( \rho'- p_r'- \frac{2}{3}\sigma' \right)+ \frac{8}{3}\sigma\left( p_r'+ \frac{2}{3}\sigma' \right) \right] ,  \label{TOV2}  \\
    \frac{d\psi}{dr} =&\ \frac{1}{\rho+ p_r}\left\lbrace -p_r'+ \alpha\left[ p_r\left( \rho'- p_r'- \frac{2}{3}\sigma' \right) + \frac{8}{3}\sigma\left( p_r'+ \frac{2}{3}\sigma' \right) \right] \left[ 16\pi+ \alpha\left(5p_r- \rho+ \frac{10}{3}\sigma\right) \right]^{-1} + \frac{2}{r}\sigma \right\rbrace ,  \label{TOV3}
  \end{align}
\end{subequations}
\end{widetext}
which will be solved numerically from the center to the surface of the star with adequate boundary conditions. Below we will give more details about this.

\subsection{Exterior solution and boundary conditions}

Here we need to impose appropriate boundary conditions for the stellar structure equations (\ref{TOV1})-(\ref{TOV3}). The numerical integration of these equations is performed from the center at $r=0$ up to some point $r= r_{\rm sur}$ where the radial pressure $p_r(r)$ vanishes. We interpret this radial coordinate as being the radius of the anisotropic compact star with central density $\rho_c$. Note that the solution of Eq.~(\ref{PsiPrime}) can be written as
\begin{align}
    \psi(r) &= -\int_r^\infty \bigg[ \frac{\alpha}{4}\left( 3p_r(\bar{r}) - \rho(\bar{r})+ \frac{2}{3}\sigma(\bar{r}) \right)\bar{r}^3p_r(\bar{r})  \nonumber \\
    &+ m(\bar{r}) + 4\pi \bar{r}^3p_r(\bar{r}) - \frac{2\alpha}{9}\bar{r}^3\sigma^2(\bar{r}) \bigg]\frac{d\bar{r}}{\bar{r}^2- 2\bar{r}m(\bar{r})} ,  \quad
\end{align}
where $\psi(\infty)= 0$. This is because spacetime is asymptotically flat at infinity. In addition, outside the star (i.e., when $r\geq r_{\rm sur}$), the quantities $\rho(r)$, $p_r(r)$ and $\sigma(r)$ vanish due to vacuum. Consequently, the last integral becomes 
\begin{equation}\label{psiOutSolution}
    \psi(r) = -\int_r^\infty\frac{Md\bar{r}}{\bar{r}^2- 2M\bar{r}} = \frac{1}{2}\ln\left[ 1- \frac{2M}{r} \right] . 
\end{equation}

Evidently, Eqs.~\eqref{ExpLambda} and \eqref{psiOutSolution} indicate that the exterior spacetime of a compact star in $f(R,T,L_m)= R+ \alpha TL_m$ gravity is still described by the familiar Schwarzschild metric, namely
\begin{equation}\label{SchwarMetric}
ds^2= -H(r)dt^2+\frac{1}{H(r)}dr^2 + r^2\left[ d\theta^2 + \sin^2\theta d\phi^2 \right] ,
\end{equation}
with
\begin{equation}
H(r)= 1- \frac{2M}{r} .
\end{equation}

We have a set of three first-order differential equations (\ref{TOV1})-(\ref{TOV3}) for the five variables $m$, $\rho$, $p_r$, $\sigma$ and $\psi$. Nonetheless, given an EoS of the form $p_r= p_r(\rho)$ and an anisotropy profile $\sigma= \sigma(m, p_r)$, we manage to reduce the number of variables to three. We will discuss these assumptions in more detail in section \ref{Sec4}. Eqs.~(\ref{TOV1}) and (\ref{TOV2}) can be integrated for a given central density and by guaranteeing regularity at the origin. Moreover, the exterior spacetime of the anisotropic fluid sphere must be described by the above Schwarzschild metric (\ref{SchwarMetric}) so that the continuity of the metric at the surface imposes a boundary condition for the other differential equation (\ref{TOV3}). In summary, the modified TOV equations (\ref{TOV1})-(\ref{TOV3}) will be solved under the fulfillment of the following boundary conditions
\begin{align}\label{BC}
    \rho(0) &= \rho_c,  &  m(0) &= 0,  &  \psi(r_{\rm sur}) &= \frac{1}{2}\ln\left[ 1- \frac{2M}{r_{\rm sur}} \right] .
\end{align}

\subsection{Moment of inertia}

In his seminal paper, Hartle \cite{Hartle1967} studied equilibrium configurations of slowly rotating stars within the framework of GR, where the moment of inertia is determined by solving a slow rotation equation. Staykov et al.~\cite{Staykov2014} followed a similar procedure in the Starobinsky gravity model. Here we will get the relativistic expression for the moment of inertia in $f(R,T,L_m)= R+ \alpha TL_m$ gravity, but also taking into consideration the tangential pressure component $p_t$. Keeping only first-order terms in the angular velocity of the stellar fluid ($\Omega$), the four-velocity is given by $u^\mu= (e^{-\psi},0,0,\Omega e^{-\psi})$ and the spacetime metric can be written in the form
\begin{align}\label{RotMetric}
  ds^2 =& -e^{2\psi(r)}dt^2 + e^{2\lambda(r)}dr^2 + r^2(d\theta^2 + \sin^2\theta d\phi^2)  \nonumber  \\
  & -2\omega(r,\theta)r^2\sin^2\theta dtd\phi ,
\end{align}
which is the slow rotation substitute for the standard line element (\ref{MetricEq}). Remark that $\psi(r)$ and $\lambda(r)$ are still functions of the radial coordinate only and $\omega(r,\theta)$ stands for the angular velocity of the inertial frames dragged by the stellar rotation \cite{Breu2016}. In what follows, we will maintain only first-order terms in $\Omega$, so that the $03$-component of the Einstein tensor for the metric (\ref{RotMetric}) assumes the form
\begin{align}
    G_{03} =&\ \frac{\sin^2\theta}{2}\frac{e^{\psi-\lambda}}{r^2}\frac{\partial}{\partial r}\left[ e^{-(\psi+\lambda)}r^4\frac{\partial\omega}{\partial r} \right]  \nonumber  \\
    &- \frac{\omega r^2\sin^2\theta}{e^{2\lambda}}\left[ \psi''+ \psi'^2 - \psi'\lambda' + \frac{1}{r}(\psi' - \lambda') \right]  \nonumber  \\
    &+ \frac{1}{2\sin\theta}\frac{\partial}{\partial\theta}\left[ \sin^3\theta\frac{\partial\omega}{\partial\theta} \right] .  \label{03ComponentEq}
\end{align}

In view of Eq.~(\ref{03ComponentEq}), the field equation (\ref{FieldEqPartForm}) yields
\begin{align}\label{03CompEq}
    &\frac{e^{\psi-\lambda}}{r^4}\frac{\partial}{\partial r}\left[ e^{-(\psi+\lambda)}r^4\frac{\partial\varpi}{\partial r} \right] + \frac{1}{r^2\sin^3\theta}\frac{\partial}{\partial\theta}\left[ \sin^3\theta\frac{\partial\varpi}{\partial\theta} \right]  \nonumber  \\
    &\hspace{0.4cm}= \left[ 16\pi+ \alpha\left( -\rho+ 5p_r+ \frac{10}{3}\sigma \right) \right](\rho+ p_t)\varpi , 
\end{align}
where $\varpi = \Omega- \omega$. Expanding $\varpi(r,\theta)$ in the form \cite{Pretel2022MPL, Hartle1967, Staykov2014}
\begin{equation}\label{ExpandEq}
    \varpi(r,\theta) = \sum_{l=1}^\infty \varpi_l(r)\left( \frac{-1}{\sin\theta}\frac{dP_l}{d\theta} \right) ,
\end{equation}
where $P_l$ are Legendre polynomials, we then obtain
\begin{align}\label{OmegaEq1}
    &\frac{e^{\psi-\lambda}}{r^4}\frac{d}{dr}\left[ e^{-(\psi+\lambda)}r^4\frac{d\varpi_l}{dr} \right] - \frac{l(l+1)-2}{r^2}\varpi_l  \nonumber  \\
    &\hspace{0.4cm}= \left[ 16\pi+ \alpha\left( -\rho+ 5p_r+ \frac{10}{3}\sigma \right) \right](\rho+ p_t) \varpi_l . 
\end{align}

At large distances from the compact star the spacetime must be asymptotically flat, so that $\varpi \rightarrow \Omega - 2J/r^3$ at large $r$, where $J$ is the total angular momentum of the star. Since the asymptotic solution of Eq.~(\ref{OmegaEq1}) is given by $\varpi_l(r) \rightarrow c_1r^{-l-2} + c_2r^{l-1}$, we can infer that $l=1$. Besides, reminding that $P_1(\cos\theta)= \cos\theta$, we identify that $\varpi$ is a function of $r$ only, and hence Eq.~(\ref{OmegaEq1}) becomes
\begin{align}\label{OmegaEq2}
    &\frac{e^{\psi-\lambda}}{r^4}\frac{d}{dr}\left[ e^{-(\psi+\lambda)}r^4\frac{d\varpi}{dr} \right]  \nonumber \\
    &\hspace{0.4cm}= \left[ 16\pi+ \alpha\left( -\rho+ 5p_r+ \frac{10}{3}\sigma \right) \right](\rho+ p_t)\varpi .
\end{align}
For an arbitrary choice of the central value $\varpi(0)$, boundary conditions for the last differential equation are imposed naturally, i.e.
\begin{align}\label{BCMomIner}
    \left. \frac{d\varpi}{dr}\right\vert_{r=0} &= 0,  &  \lim_{r\rightarrow\infty}\varpi &= \Omega ,
\end{align}
where the first ensures regularity at the center of the star and the second satisfies the asymptotic flatness requirement at infinity. Of course, to ensure the continuity and differentiability of the solution, one has to solve the problem inside and outside the stellar fluid by means of the following matching conditions
\begin{align}
    \varpi_{in}(r_{\rm sur}) &= \varpi_{out}(r_{\rm sur}),  \\
    \left. \frac{d\varpi_{in}}{dr}\right\vert_{r= r_{\rm sur}} &= \left. \frac{d\varpi_{out}}{dr}\right\vert_{r= r_{\rm sur}} .
\end{align}

Finally, through slow rotation equation (\ref{OmegaEq2}) and the asymptotic form of $\varpi(r)$ we can get an explicit expression for the relativistic moment of inertia of slowly rotating anisotropic compact stars in $f(R,T,L_m)= R+ \alpha TL_m$ gravity:
\begin{widetext}
  \begin{equation}\label{MomInerEq}
  I = \int_0^{r_{\rm sur}} \left[ \frac{8\pi}{3}+ \frac{\alpha}{6}\left( -\rho+ 5p_r+ \frac{10}{3}\sigma \right) \right] \left[1+ \frac{\sigma}{\rho+ p_r}\right](\rho+ p_r) e^{\lambda-\psi}r^4 \left( \frac{\varpi}{\Omega} \right)dr ,
  \end{equation}
\end{widetext}
which reduces to the GR expression for anisotropic compact stars when $\alpha =0$ \cite{Pretel2024PLB}. In addition, when both $\alpha$ and $\sigma$ vanish, we recover the conventional moment of inertia for isotropic fluid configurations in pure Einstein gravity \cite{Glendenning}. It is important to highlight that, for the metric (\ref{RotMetric}), the stellar structure equations (\ref{TOV1})-(\ref{TOV3}) remain the same to rotational corrections at first order in the angular velocity $\Omega$. In other words, similar to the standard GR context, the spherical symmetry is still preserved but an additional differential equation arises from the $03$-component of the field equations, which has allowed us to obtain the moment of inertia in the slowly rotating approximation.


\section{Equation of state and anisotropy profile}\label{Sec4}

Describing isotropic compact stars in any gravity theory involves providing a crucial input known as EoS, which describes the dense microscopic matter that makes up the star. At asymptotically high densities, the most favourable state for strange quark matter is the color-flavour-locked
(CFL) phase \cite{Oikonomou2023}. In the present study therefore we will focus on QSs composed of color superconducting quark matter. Below we will briefly describe the relation between radial pressure and energy density for such stars.

To order $\Delta^2$ including perturbative QCD corrections, where $\Delta$ is the gap energy which accounts for quark interactions, the thermodynamic potential of the CFL phase in dense matter is given by \cite{Alford2001, Lugones2002}
\begin{align}\label{potentialCFL}
    \Omega =&\ \frac{6}{\pi^2}\int_0^\nu \texttt{p}^2(\texttt{p}- \mu)d\texttt{p} + \frac{3}{\pi^2}\int_0^\nu \texttt{p}^2\left( \sqrt{\texttt{p}^2+ m_s^2} - \mu \right)d\texttt{p}  \nonumber  \\
    &+ \frac{3\mu^2}{4\pi^2}(1- a_4) - \frac{3\Delta^2\mu^2}{\pi^2} + B , 
\end{align}
where $3\mu= \mu_u+ \mu_d+ \mu_s$ is the baryon chemical potential, $m_s$ is the strange quark mass, and the pQCD corrections are represented by the $(1-a_4)$ term up to $\mathcal{O}(a_s^2)$, with $a_4$ being a quartic coefficient that varies from $a_4= 1$ (when strong interactions are not considered) to rather small values when these interactions become important. Notice that $B$ is the effective bag constant which depicts the nonperturbative QCD
vacuum effects. Moreover, the common Fermi momentum takes the form 
\begin{equation}
    \nu = 2\mu- \sqrt{\mu^2+ \frac{m_s^2}{3}} .
\end{equation}

By expanding the thermodynamic potential in powers of $m_s/\mu$, Eq.~(\ref{potentialCFL}) leads to the EoS of strange stars composed of CFL quark matter \cite{Oikonomou2023}:
\begin{equation}\label{EqoS}
    \rho= 3p_r+ 4B + \frac{3w^2}{4\pi^2} + \frac{w}{\pi}\sqrt{3(p_r+ B) + \frac{9w^2}{16\pi^2}} ,
\end{equation}
where the following parameters have been defined
\begin{align}
    k &= \sqrt{a_4},  &  a_2 &= m_s^2- 4\Delta^2,  &  w &= \frac{a_2}{k} .
\end{align}

We must point out that the term associated with the gap parameter $\Delta$ in Eq.~(\ref{potentialCFL}) describes the color superconductivity contribution. Following Ref.~\cite{Oikonomou2023}, in all our numerical calculations we will use $m_s= 95\, \rm MeV$ and $a_4= 0.65$. Furthermore, we will choose two different $\Delta-B^{1/4}$ pairs, namely $\bigl\{ \Delta= 80\, {\rm MeV}, B^{1/4}= 130.6\, \rm MeV \bigr\}$ and $\bigl\{\Delta= 180\, {\rm MeV}, B^{1/4}= 161\, \rm MeV \bigr\}$, referred to as EoS I and EoS II, respectively. The reason for such choices is because they allow us to obtain color-superconducting QSs in GR that consistently satisfy the observational constraints, see Ref.~\cite{Oikonomou2023} for further details.

The absolute stability of strange quark matter (SQM) has been proposed in Ref.~\cite{Farhi1984}, requiring the minimum value of the energy per baryon of SQM should be less than the minimum energy per baryon of the observed most stable nuclei $M(\rm ^{56}Fe)/56$, i.e.~$930\, \rm MeV$. Thus, the energy per baryon number $E/A$ is crucial for the stability examination of CFL quark
matter, given by \cite{Oikonomou2023}
\begin{equation}\label{EnerBaryNEq}
    \left( \frac{E}{A} \right)_{\rm CFL} = \frac{2\sqrt{6}\pi}{k^{1/2}} \frac{B^{1/2}}{\sqrt{\sqrt{\frac{16\pi^2}{3}B + w^2} - w}} ,
\end{equation}
which is a result similar to the one derived in Ref.~\cite{Zhang2021}. For the above equations of state I and II we have that the minimum energy per baryon is $765$ and $712\, \rm MeV$, respectively, which are less than $930\, \rm MeV$.

If the stellar fluid were isotropic, an EoS would be sufficient to close the system of stellar structure equations. Nevertheless, our work aims to cover more general astrophysical scenarios, where in addition to a radial pressure there is a tangential pressure. Thus, the need arises to implement an anisotropy profile that allows us to describe anisotropies within the QS. Of course, under a certain limit we must recover the isotropic solutions. With this in mind, we consider the anisotropy model suggested by Horvat and collaborators, where the transverse pressure is given by \cite{Horvat2011}
\begin{equation}
    p_t = p_r\left[ 1+ \beta(1-e^{-2\lambda}) \right] ,
\end{equation}
or alternatively,
\begin{equation}\label{AniProfileEq}
    \sigma = \beta\left( \frac{2m}{r} \right)p_r ,
\end{equation}
with $\beta$ being a free parameter that determines the degree of anisotropy within the compact star. This profile obeys a set of physical acceptability conditions and, in order not to repeat them, we refer the reader to Refs.~\cite{Horvat2011, Yagi2015, Folomeev2018, Pretel2020EPJC, Rahmansyah2020, Rahmansyah2021} for more detailed discussions. As is evident, the stellar fluid becomes isotropic when $\beta= 0$.

\section{Numerical results}\label{ResultsSec}

\subsection{Isotropic configurations}

\begin{figure*}
    \centering
        \includegraphics[width=8.5cm]{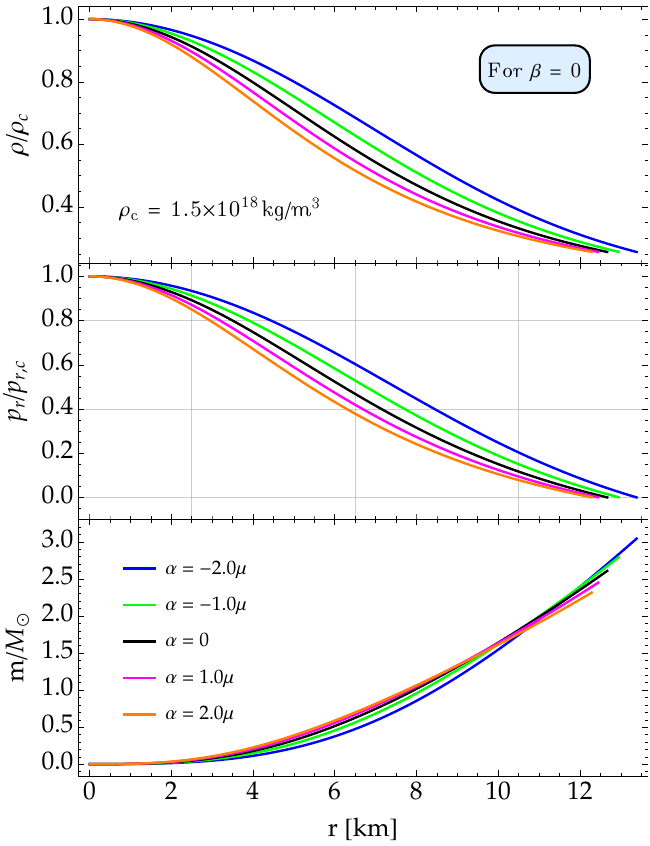}
        \includegraphics[width=8.5cm]{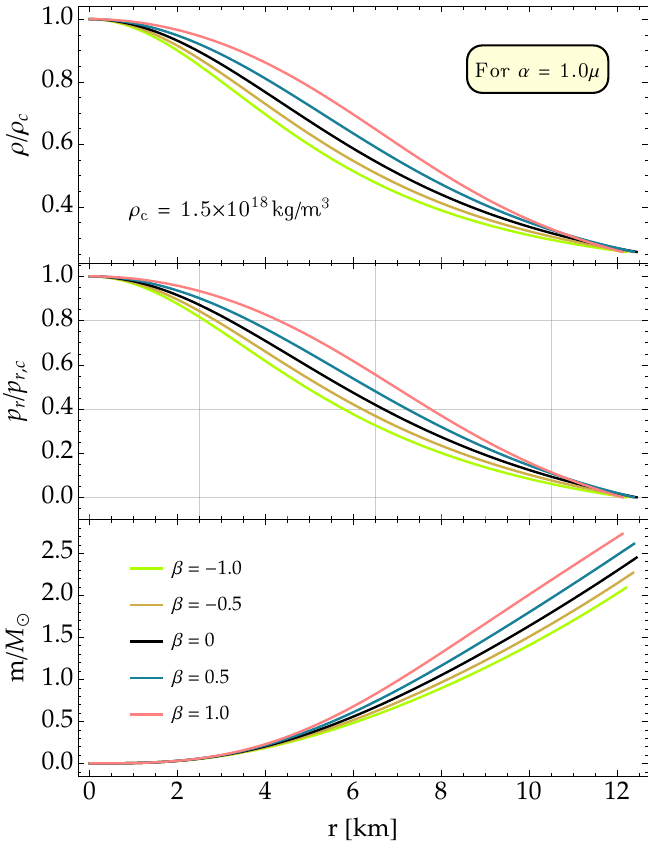}
       \caption{Normalized energy density, radial pressure and mass distribution as functions of the radial coordinate $r$ for a given central density $\rho_c= 1.5 \times 10^{18}\, \rm kg/m^3$ and CFL quark matter with EoS II. In the left plot we have isotropic solutions (this is, when $p_t= 0$) varying the coupling constant in the range $\alpha \in [-2.0, 2.0]\mu$, where $\mu= 10^{10}\, \rm m^2$ and the particular case $\alpha= 0$ corresponds to the pure GR solution. In the right plot we have considered five values for the anisotropy parameter (with $\vert\beta\vert \leq 1.0$ and $\beta= 0$ representing the isotropic solution) and fixed $\alpha= 1.0\mu$. }
    \label{FigRadialBehavior}
\end{figure*}

\begin{figure*}
    \centering
        \includegraphics[width=8.804cm]{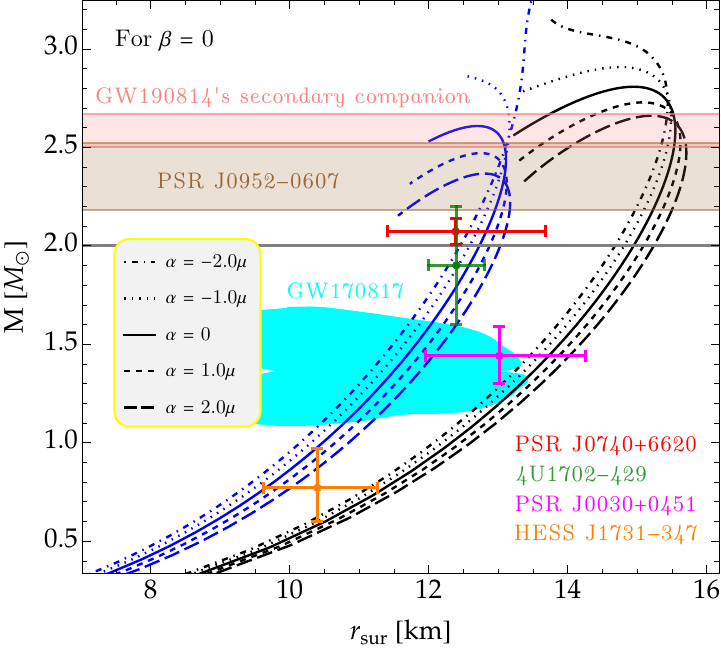}
        \includegraphics[width=8.75cm]{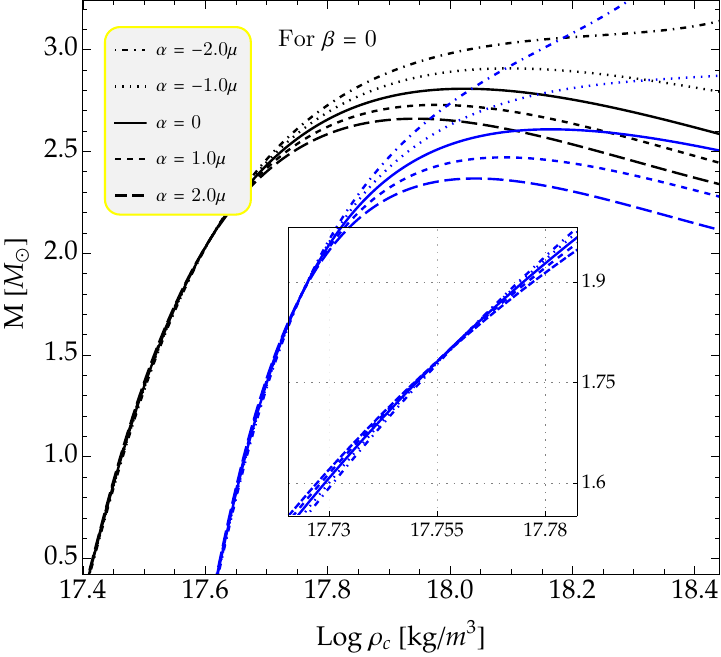}
       \caption{Mass-radius diagrams (left panel) and mass-central density relations (right panel) for isotropic QSs in $f(R,T,L_m)= R+ \alpha TL_m$ gravity for five values of $\alpha$. These stellar configurations correspond to $\sigma= 0$ (i.e., when $\beta= 0$) and CFL EoS (\ref{EqoS}) for two sets of $\Delta-B^{1/4}$ pairs: $\bigl\{ \Delta= 80\, {\rm MeV}, B^{1/4}= 130.6\, \rm MeV \bigr\}$ by black lines and $\bigl\{\Delta= 180\, {\rm MeV}, B^{1/4}= 161\, \rm MeV \bigr\}$ by blue curves. The gray horizontal streak at $2M_\odot$ represents the two massive NS pulsars J1614-2230 \cite{Demorest2010} and J0348+0432 \cite{Antoniadis2013}. The filled brown and pink bands stand for the masses of the fastest known spinning NS in the disk of the Milky Way (namely, the pulsar PSR J0952-0607 \cite{Romani2022}) and of the secondary companion detected by the gravitational-wave signal GW190814 \cite{Abbott2020}, respectively. The cyan region is the mass-radius constraint obtained from the GW170817 event \cite{Abbott2018PRL}. The red, green, magenta and orange dots with their respective error bars represent the masses of massive millisecond pulsars PSR J0740+6620 \cite{Riley2021}, 4U 1702-429 \cite{Nattila2017}, PSR J0030+0451 \cite{Miller2019} and supernova remnant HESS J1731-347 \cite{Doroshenko2022}, respectively. }
    \label{FigMRd}
\end{figure*}

\begin{figure*}
    \centering
        \includegraphics[width=8.65cm]{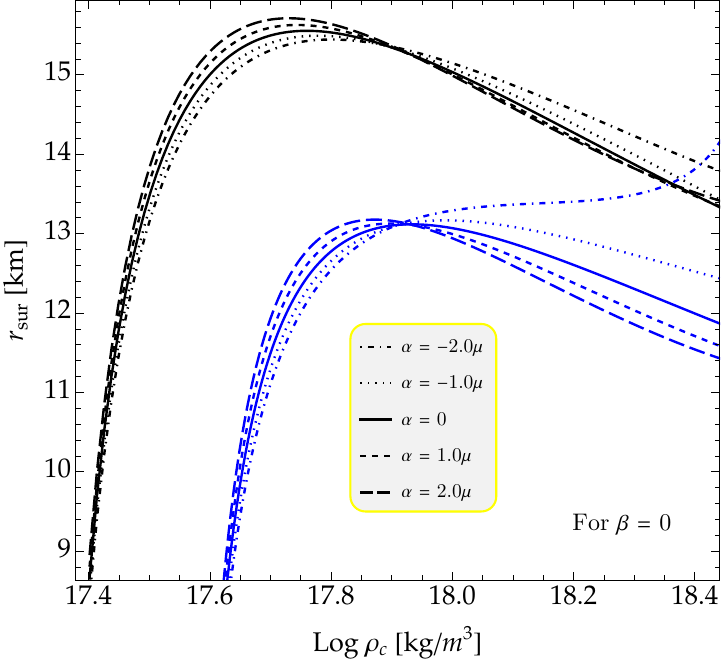}
        \includegraphics[width=8.73cm]{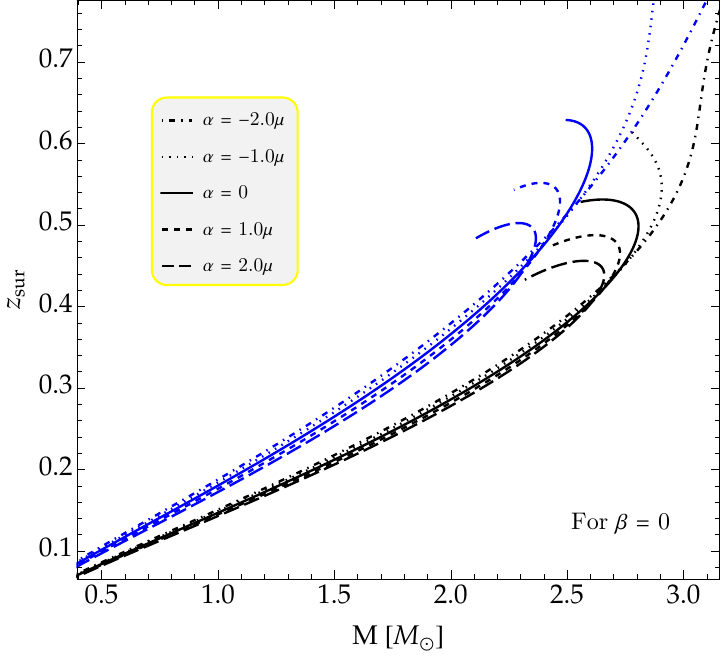}
       \caption{Left-hand panel: Surface radius as a function of the central density for the isotropic stellar configurations shown in Fig.~\ref{FigMRd}. Right-hand panel: Surface gravitational redshift versus total mass. The solid curves correspond to the pure general relativistic solutions (that is, when $\alpha= 0$). }
    \label{FigRdzM}
\end{figure*}

We begin our analysis by examining isotropic QSs, i.e., when $\sigma= 0$ in all stellar structure equations. With the specification of an EoS, the set of differential equations (\ref{TOV1})-(\ref{TOV3}) is solved simultaneously from the center of the star at $r=0$, where we give as input a certain value of $\rho_c$, to its surface at $r= r_{\rm sur}$. In the left plot of Fig.~\ref{FigRadialBehavior} we show the radial behavior of the energy density, pressure and mass function for isotropic quark stars with $\rho_c= 1.5 \times 10^{18}\, \rm kg/m^3$ in $f(R,T,L_m)= R+\alpha TL_m$ gravity for five values of $\alpha$, where the black curve indicates the pure general relativistic solution (i.e., when $\alpha= 0$). Note that the parameter $\alpha$ is given in $\mu= 10^{10}\, \rm m^2$ units. We observe that both energy density and pressure are decreasing functions with the radial coordinate, while mass increases as we approach the surface. As expected, the pressure vanishes at the surface and the energy density $\rho(r_{\rm sur})$ assumes a non-zero value due to the adopted EoS. Note further that the impact of the parameter $\alpha$ on the relativistic structure of the compact star is an increase in both radius $r_{\rm sur}$ and total mass $M$ as $\alpha$ becomes more negative. However, the opposite occurs when $\alpha$ assumes positive values. We will see later that this qualitative behavior is not always maintained for the entire range of central energy densities. In other words, such behavior can be reversed in some other band of $\rho_c$ (see the blue curves in the left panel of Fig.~\ref{FigRdzM} for central densities smaller than the value considered above), which ends up being a peculiar characteristic of the gravity model considered in this work.

Sampling the values $\rho_c$, a sequence of equilibrium configurations can be constructed for different values of $\alpha$, which are represented in the well-known mass-radius diagram, see the left plot of Fig.~\ref{FigMRd}. For both sets of QSs (namely, blue and black curves for two sets of $\Delta-B^{1/4}$ pairs), the main consequence of the parameter $\alpha$ is to substantially modify the $M-r_{\rm sur}$ relation. Its influence is most notable in the maximum-mass values: Positive $\alpha$ decreases the maximum mass $M_{\rm max}$, while negative values lead to an increase in $M_{\rm max}$. As an atypical result of gravity theory, for sufficiently large values of $\vert\alpha\vert$ keeping negative $\alpha$, the critical point where the mass is maximum cannot be found on the mass-radius diagram. This turns out to be very interesting because perhaps the stars never stop being stable for very negative values of $\alpha$. Of course, a more rigorous analysis on the stability of these stars would involve considering small radial perturbations around the equilibrium state, as done in Einstein gravity. Due to the extensive and tedious treatment of the perturbed field equations, we will leave this analysis for a future study.

It is expected that our theoretical predictions of the $f(R,T,L_m)$ gravity model are capable of describing compact stars observed in the universe, e.g., millisecond pulsars \cite{Demorest2010, Antoniadis2013, Riley2021, Nattila2017, Miller2019, Romani2022} and other compact objects which are still of unknown nature \cite{Abbott2020, Doroshenko2022}. In fact, different astrophysical measurements, based on X-ray emissions and on gravitational wave observations, play an important role in adopting or discarding a gravity theory. Therefore, it is essential to compare our results with recent observational measurements/constraints. For both EoSs I (black) and II (blue lines) in the left panel of Fig.~\ref{FigMRd}, our theoretical predictions consistently describe the pulsar PSR J0952-0607, the fastest known spinning NS in the disk of the Milky Way \cite{Romani2022}. For the range $\vert\alpha\vert \leq 2.0\mu$ and using EoS II, we can observe that the Bayesian estimations for the massive pulsar PSR J0740+6620 \cite{Riley2021} and the NS in 4U 1702-429 \cite{Nattila2017} are in agreement with our numerical results. Note also that, at small masses and EoS II, positive values of $\alpha$ favor the description of the central compact object within the supernova remnant HESS J1731-347 \cite{Doroshenko2022}. On the other hand, our results for EoS I (black curves) are consistent with the estimates of mass and radius of the millisecond pulsar PSR J0030+0451 \cite{Miller2019}.

Now let us analyze the relation between mass and central density for both EoSs, see the right plot of Fig.~\ref{FigMRd}. For low central densities, the total mass of the star decreases (increases) slightly for negative (positive) values of $\alpha$, see panel inserted in the right plot of Fig.~\ref{FigMRd}. Notwithstanding, after a certain value of $\rho_c$, this behavior is reversed and much more pronounced. Therefore, regardless of the CFL EoS, the alterations in the gravitational mass attributed to the $\alpha TL_m$ term are imperceptible at small $\rho_c$, while the most relevant changes take place in the high-central-density branch.

The effect of $\alpha$ on the radius can be better examined in a plot of $r_{\rm sur}$ versus $\rho_c$ as in the left panel of Fig.~\ref{FigRdzM}. The radius of the star decreases noticeably as $\alpha$ becomes more negative in the small-central-density region, however, this behavior is reversed above a certain central density value. In other words, the radius undergoes a relevant increase (decrease) with respect to its GR counterpart as $\alpha$ is more negative (positive) for large values of $\rho_c$. These results differ notably from the theoretical predictions in other gravitational contexts, such as the modification of the radius by varying $\beta$ in the $f(R,T)= R+2\beta T$ gravity model \cite{Pretel2021JCAPa}. Furthermore, the right-hand panel of Fig.~\ref{FigRdzM} displays the surface gravitational redshift, given by $z_{\rm sur}= e^{-\psi(r_{\rm sur})} - 1$, as a function of total mass. $z_{\rm sur}$ is strongly affected by the extra $\alpha TL_m$ term in the high-mass branch, while the changes are smaller at sufficiently low masses.

\begin{figure}
    \centering
        \includegraphics[width=8.6cm]{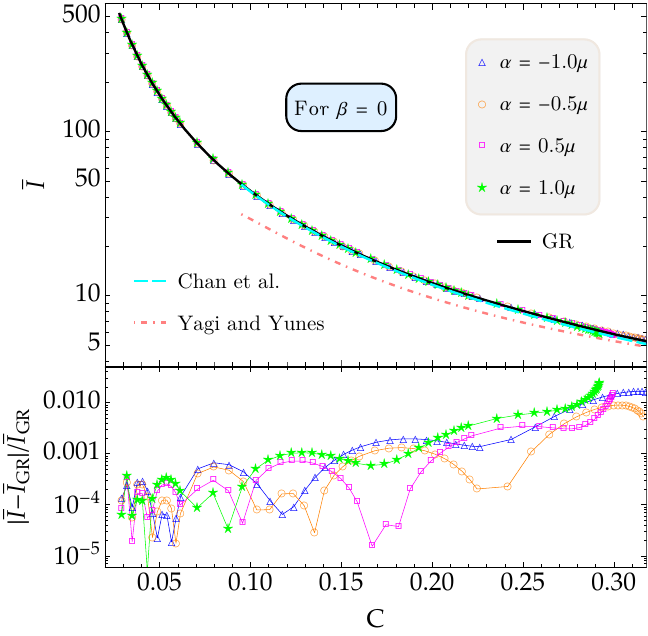}
       \caption{Normalized moment of inertia $\bar{I}$ as a function of the stellar compactness $C$ for our adopted EoS II, where the black solid curve denotes the fitted function in GR (\ref{FitEq1}) with its fitting coefficients given in Table \ref{table1}. Note also that here we have assumed $\beta= 0$. The lower panel shows the relative fractional difference between the numerical results in $f(R,T,L_m)= R+ \alpha TL_m$ and the fitting curve in GR. Moreover, the cyan dashed and pink dot-dashed curves correspond to the fits obtained for QSs and NSs within the context of Einstein gravity in Refs.~\cite{Chan2016} and \cite{YagyYunes2017}, respectively. }
    \label{FigICUniRelatBe0}
\end{figure}

An additional task of our work is to investigate whether the $I-C$ universal relation remains valid in $f(R,T,L_m)$ gravity. Namely, we will keep the EoS fixed and vary the free parameter $\alpha$ for a subsequent analysis of the deviations introduced by the $\alpha TL_m$ term with respect to the results predicted by Einstein gravity. In the top panel of Fig.~\ref{FigICUniRelatBe0}, we plot the dimensionless moment of inertia $(\bar{I}= I/M^3)$ against the stellar compactness $C$ using EoS II, where we have considered four non-null values of $\alpha$ and the black solid curve indicates the pure GR case. Inspired by previous works \cite{Chan2016, Breu2016, YagyYunes2017, 2024arXiv240112519P}, such a solid curve can be fitted through a series expansion method of the type
\begin{equation}\label{FitEq1}
    \log_{10}\bar{I} = \sum_{n=-4}^5 a_n C^n , 
\end{equation}
where $a_n$ are fitting coefficients and are summarized in Table \ref{table1}. The relative fractional difference between the numerical results obtained for $\alpha \neq 0$ and the analytical expression in GR (\ref{FitEq1}) is shown in the lower panel of Fig.~\ref{FigICUniRelatBe0}. From this figure it can clearly be observed that the $\bar{I}-C$ relations are essentially independent of the coupling parameter $\alpha$. This relation suffers a maximum deviation of approximately $3\%$ from the GR counterpart for $\alpha= 1.0\mu$, while the deviations are much smaller for relatively small compactnesses. This means that the suitably scaled moment of inertia and compactness are still connected by a robust universal relation in $f(R,T,L_m)= R+ \alpha TL_m$ gravity, which is insensitive to the values of $\alpha$ to within about $3\%$ level. Therefore, our results show that the $\bar{I}-C$ relation also holds in $f(R,T,L_m)$ gravity up to $3\%$ compared to GR predictions for the range $\vert\alpha\vert \leq 1.0\mu$.

\begin{table*}
\caption{\label{table1} Fitting coefficients for the empirical formula (\ref{FitEq1}) in the $\bar{I}-C$ relation of CFL quarks stars with EoS II, where we have assumed $\beta= 0$. The cases $\alpha= 0$ and $1.0\mu$ are represented by the black curves in Figs.~\ref{FigICUniRelatBe0} and \ref{FigICUniRelatAl1}, respectively. }
\begin{ruledtabular}
\begin{tabular}{c|cccccccccc}
Theory  &  $a_{-4} [10^{-6}]$  &  $a_{-3} [10^{-4}]$  &  $a_{-2} [10^{-2}]$  &  $a_{-1} [10^{-1}]$  &  $a_0$  &  $a_1 [10^2]$  &  $a_2 [10^2]$  &  $a_3 [10^3]$  &  $a_4 [10^3]$  &  $a_5 [10^3]$  \\
\hline
  $\alpha =0$ (GR)  &  $-2.2649$  &  $2.9593$  &  $-1.6107$  & 
 $4.8861$  &  $-5.5678$  &  $0.6787$  &  $-4.5293$  &  $1.7155$  &  $-3.4536$  &  $2.8517$  \\
  $\alpha =1.0\mu$  &  $-3.2240$  &  $4.2319$  &  $-2.3099$  &  $6.9601$  &  $-9.2072$  &  $1.0689$  &  $-7.0839$  &  $2.7012$  &  $-5.4945$  &  $4.5835$  \\
\end{tabular}
\end{ruledtabular}
\end{table*}

\subsection{Anisotropic configurations}

\begin{figure*}
    \centering
        \includegraphics[width=8.82cm]{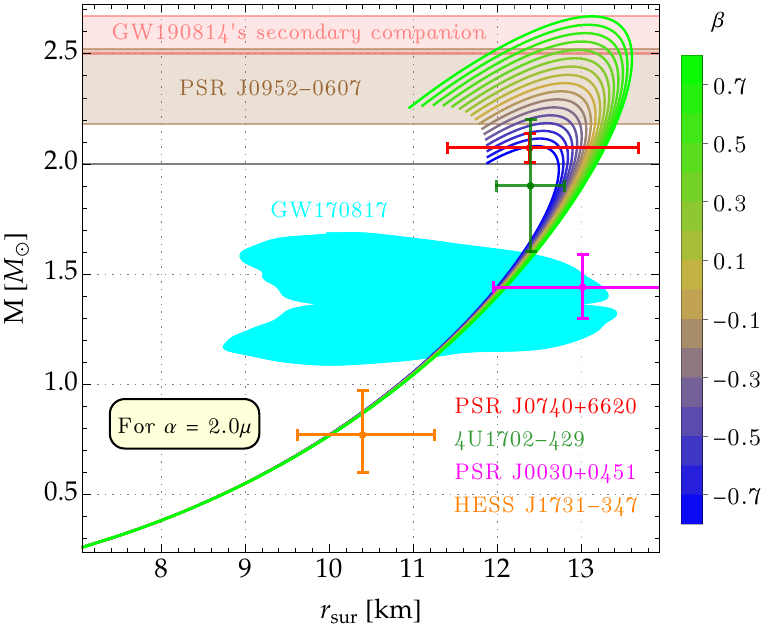}
        \includegraphics[width=8.765cm]{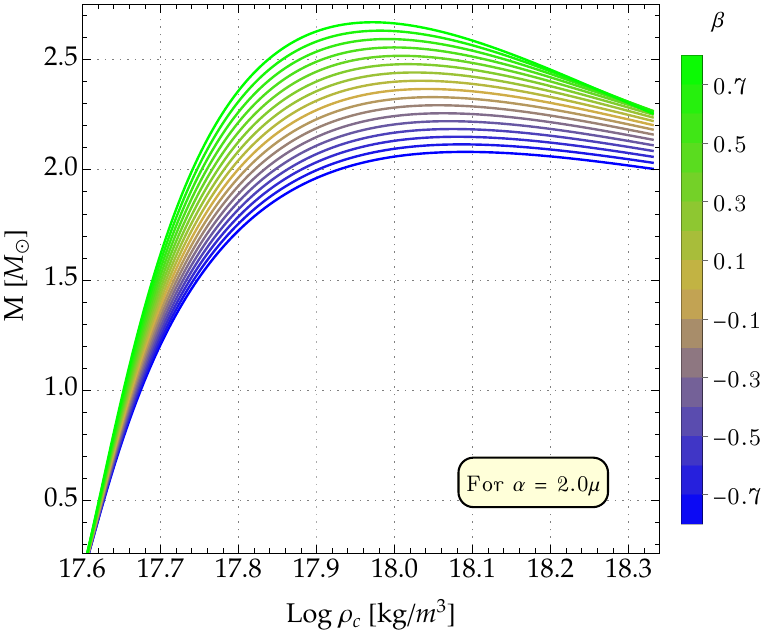}
       \caption{Anisotropic QSs composed of color superconducting quark matter in $f(R,T,L_m)$ gravity for several values of the anisotropy parameter $\beta \in[-0.8, 0.8]$ and fixed $\alpha= 2.0\mu$, where we have assumed the EoS II. The different regions and colored bars in the mass-radius diagram (left panel) represent the same observational measurements as those in the left plot of Fig.~\ref{FigMRd}. Meanwhile the right panel illustrates the mass as a function of the central density, indicating that the greatest impact of anisotropy on the mass $M$ takes place at high central densities.  }
    \label{FigMRdAniCase}
\end{figure*}

\begin{figure*}
    \centering
        \includegraphics[width=8.78cm]{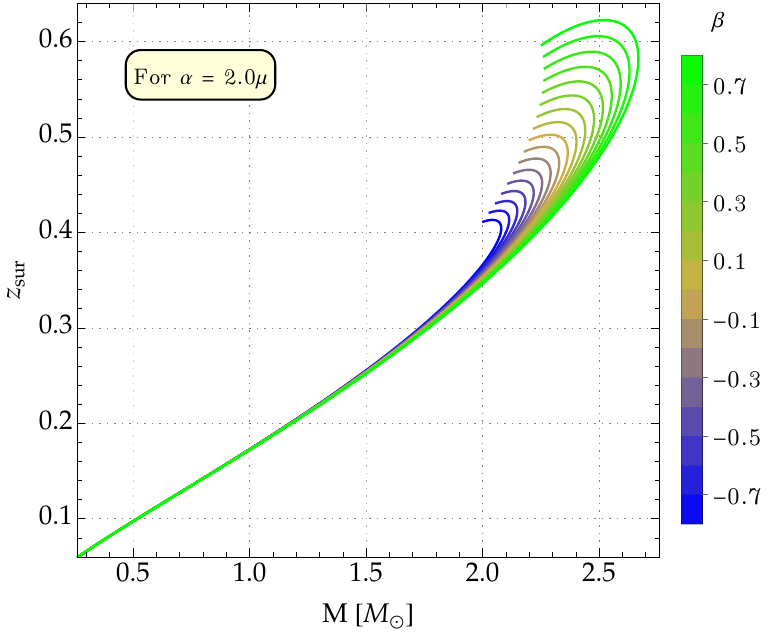}
        \includegraphics[width=8.65cm]{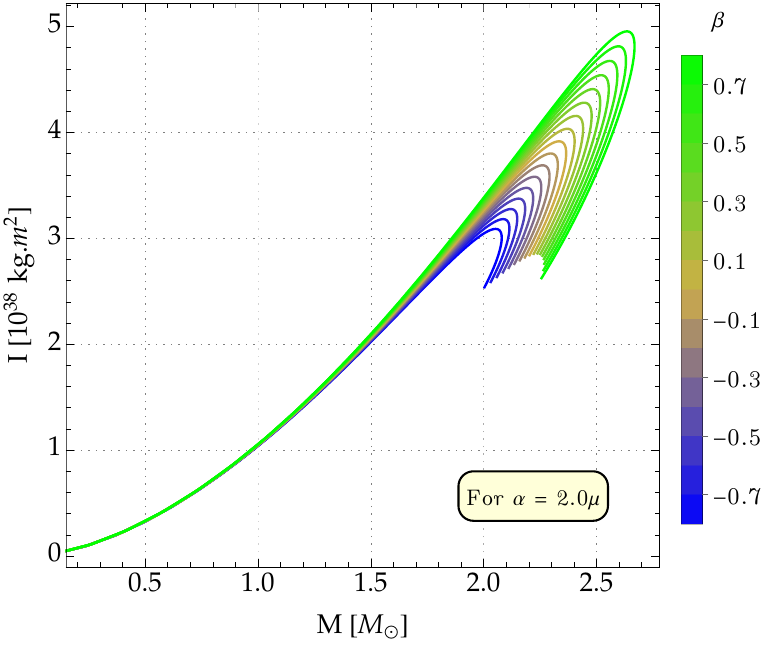}
       \caption{Surface gravitational redshift (left plot) and moment of inertia (right plot) as functions of total mass for the anisotropic CFL QSs shown in Fig.~\ref{FigMRdAniCase}. The biggest effect of anisotropic pressure on both quantities occurs predominantly in the high-mass region, while its influence is irrelevant at small masses. }
    \label{FigzIMAniCase}
\end{figure*}

As mentioned in the introduction, anisotropic pressure plays an important role in constructing more realistic scenarios of compact stars. With this in mind, here we extend our analysis of compact stars in $f(R,T,L_m)$ gravity to an anisotropic context. Similar to the isotropic case, we numerically solve the hydrostatic equilibrium equations (\ref{TOV1})-(\ref{TOV3}) for a specific value of $\rho_c$ and $\alpha$, but taking into account the anisotropy profile (\ref{AniProfileEq}). For example, for $\rho_c= 1.5 \times 10^{18}\, \rm kg/m^3$ and $\alpha = 1.0\mu$, the right plot of Fig.~\ref{FigRadialBehavior} illustrates the radial behavior of the energy density, radial pressure and mass distribution for five values of the anisotropy parameter $\beta$, where it can be seen that the internal structure of a QS in $f(R,T,L_m)= R+ \alpha TL_m $ gravity can be appreciably affected by the presence of anisotropy.

Varying the value of $\rho_c$ and keeping fixed $\alpha= 2.0\mu$ for the EoS II, we generate the mass-radius diagrams and the mass-central density relations for anisotropic QSs when the anisotropy parameter varies in the range $\vert \beta\vert \leq 0.8$, as shown in Fig.~\ref{FigMRdAniCase}. Similar to the results reported in pure Einstein gravity \cite{Curi2022, Arbanil2023, Pretel2024PLB}, it can be observed that the substantial impact on quark star structure due to the presence of anisotropy only takes place in the high-mass branch. In effect, the maximum mass can increase (decrease) significantly for positive (negative) values of $\beta$, which favors the description of highly massive compact objects such as the secondary component detected from the binary coalescence GW190814 \cite{Abbott2020}, see green curves when $0.4 \lesssim \beta \lesssim 0.8$. Remarkably, our theoretical predictions for $M-r_{\rm sur}$ relations of anisotropic QSs in $f(R,T,L_m)= R+ \alpha TL_m $ gravity consistently describe the observational measurements of different massive millisecond pulsars \cite{Demorest2010, Antoniadis2013, Riley2021, Nattila2017, Miller2019, Romani2022} as well as the central compact object within the supernova remnant HESS J1731-347 \cite{Doroshenko2022}. Consequently, anisotropic pressure effects also play a crucial role in $f(R,T,L_m)$ modified gravity theories.

According to the left panel of Fig.~\ref{FigzIMAniCase}, the gravitational redshift of light emitted at the surface of a QS is significantly affected by $\beta$ for high masses, while its impact is irrelevant in the small-mass region. By means of the integral (\ref{MomInerEq}), we have determined the moment of inertia for the anisotropic QS sequences shown in Fig.~\ref{FigMRdAniCase}. Note that before carrying out such an integration it is necessary to solve the differential equation (\ref{OmegaEq2}) inside and outside the star with the respective boundary conditions (\ref{BCMomIner}). The right plot of Fig.~\ref{FigzIMAniCase} illustrates the moment of inertia as a function of gravitational mass, and we can appreciate that the most important consequence of a positive anisotropy is to increase the maximum values of $I$, while negative anisotropies have the opposite effect. Therefore, this qualitative behavior, already predicted by conventional GR, is maintained in the $f(R,T,L_m)= R+ \alpha TL_m$ gravity model.

\begin{figure}
    \centering
        \includegraphics[width=8.6cm]{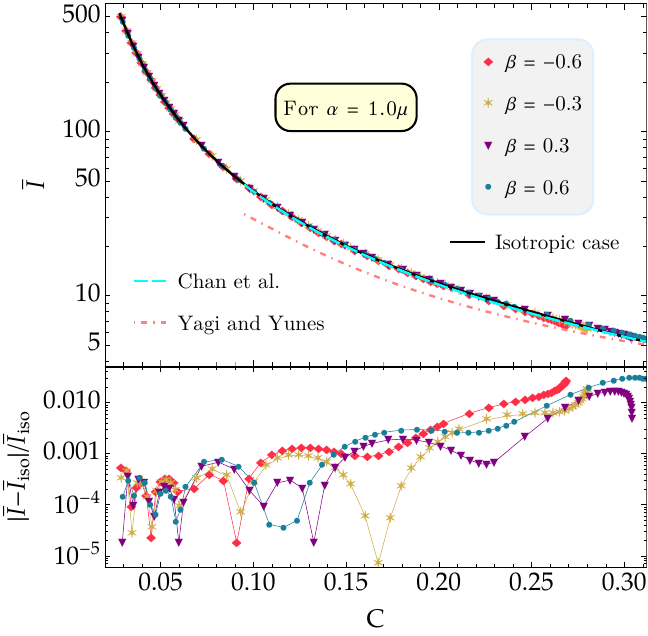}
       \caption{Upper panel: The dimensionless moment of inertia $\bar{I}$ is plotted against compactness $C$ for different values of the anisotropy parameter $\beta$ and fixed $\alpha= 1.0\mu$. The EoS is the same used in Fig.~\ref{FigICUniRelatBe0}. The black solid line is the fitting curve for the data extracted assuming $\beta=0$, i.e.~the isotropic case. The fitting parameters $a_n$ for such a curve are given in the last row of Table \ref{table1}. Lower panel: Relative fractional difference between the numerical results for $\beta \neq 0$ and the fitting curve. }
    \label{FigICUniRelatAl1}
\end{figure}

Finally, we will explore whether the relation between the normalized moment of inertia $\bar{I}$ and compactness $C$ is independent of the anisotropy parameter $\beta$. For fixed $\alpha= 1.0\mu$, Fig.~\ref{FigICUniRelatAl1} shows this relation for four values of $\beta$, and the black solid curve represents the isotropic case. The fitting parameters $a_n$ are shown in the last row of Table \ref{table1}. According to the lower panel, the relative difference $\Delta I= \vert \bar{I} - \bar{I}_{\rm Iso} \vert/\bar{I}_{\rm Iso}$ indicates that $\bar{I}$ can undergoes variations with respect to its isotropic counterpart of up to $\sim 4\%$. Our results therefore reveal that the $\bar{I}-C$ correlation for QSs is insensitive to the presence of anisotropy not only in GR (as shown by previous studies \cite{2024arXiv240112519P}) but also in the context of $f(R,T,L_m)$ gravity.

\section{Final remarks}\label{Conclusions}

Throughout the present paper we focused our attention on astrophysical implications of $f(R,T,L_m)$ gravity on compact stars. For this purpose, we have used the simplest gravity model given by $f(R,T,L_m)= R+ \alpha TL_m$, where $\alpha$ is a free parameter of the theory. Under such a gravitational context and considering an anisotropic perfect fluid, we derived for the first time the modified TOV equations and the relativistic moment of inertia in the slowly rotating approximation, where only first-order rotational corrections appear in the angular velocity of the star. This work has been approached in such a way that we have answered the question: How the relativistic structure of a compact star is affected both by the modification of Einstein's theory of gravity via the $\alpha TL_m$ term and by the presence of anisotropy?

As a first step, we have numerically solved the stellar structure equations for isotropic color-superconducting QSs in order to examine the influence of $\alpha$ on the $M-r_{\rm sur}$ diagrams. We found that the decrease of $\alpha$ significantly increases the maximum-mass values. We obtained a qualitative behavior similar to the pure Einstein gravity for negative $\alpha$ with sufficiently small $\vert\alpha\vert$, i.e., it is possible to get a critical stellar configuration such that the mass reaches its maximum. Nonetheless, for sufficiently large values of $\vert\alpha\vert$ keeping negative $\alpha$, the critical point cannot be found on the $M-r_{\rm sur}$ diagram. Additionally, the radius undergoes a relevant increase (decrease) with respect to its GR counterpart as $\alpha$ becomes more negative (positive) in the high-central-density region. Meanwhile, for small values of $\rho_c$, this behavior is opposite as $\alpha$ varies. For very small central densities, the alterations attributed to the coupling parameter $\alpha$ on radius and mass are negligible. Consequently, we can conclude that this gravity model differs significantly from the pure GR context only at high central energy densities.

Since anisotropy plays a crucial role in describing self-gravitating objects, our second goal was to examine the effect of anisotropic pressure on the relativistic structure of QSs within the framework of $f(R,T,L_m)$ gravity theories. Similar to the results reported in GR, our findings revealed that the main consequence of introducing positive (negative) anisotropy is a significant increase (decrease) in the maximum values of mass, gravitational redshift and moment of inertia for a fixed $\alpha$. In addition, our theoretical mass-radius predictions of anisotropic QSs in $f(R,T,L_m)= R+ \alpha TL_m$ gravity outstandingly favor the observational measurements of different massive millisecond pulsars as well as the central compact object within the supernova remnant HESS J1731-347.

We have also studied the $I-C$ correlation for color-superconducting QSs in $f(R,T,L_m)$ gravity in the slow-rotation approximation. Specifically, for $\vert\alpha\vert \leq 1.0\mu$, we found that the $I-C$ universal relation discovered in Einstein gravity remains valid in $f(R,T,L_m)= R+ \alpha TL_m$ gravity model, at least to $3\%$ compared to the general relativistic predictions. In similar manner, for a fixed value of $\alpha$, our outcomes show that the $I-C$ relation is independent of the degree of anisotropy within the QS. Specifically, we found that this relation is insensitive to variations of $\beta$ to $\mathcal{O}(4\%)$.

Overall, this work discusses the importance of understanding the basic macroscopic properties of quark stars in $f(R,T,L_m)$ modified gravity theories, not only including isotropic but also anisotropic matter in order to cover more realistic scenarios. In addition to quark matter, it is also worth considering hadronic matter models as well as other anisotropy profiles. In the future therefore this study may be extended to consider a variety of EoSs and verify the validity of other universal relations involving tidal deformability and $f$-mode frequency.

\begin{acknowledgments}
JMZP acknowledges support from ``Fundação Carlos Chagas Filho de Amparo à Pesquisa do Estado do Rio de Janeiro'' -- FAPERJ, Process SEI-260003/000308/2024.
\end{acknowledgments}

\end{document}